\documentclass[aps, prd, 12pt, eqsecnum, tightenlines, notitlepage, superscriptaddress, nofootinbib, floatfix]{revtex4-1}

\PassOptionsToPackage{unicode}{hyperref}

\usepackage{hyperref, graphicx, multirow, slashed, xcolor, amsmath, breakurl}

\allowdisplaybreaks

\definecolor{red}{rgb}{1,0,0}

\usepackage{braket}

\begin{document}

\title{An Estimate of the Inclusive Branching Ratio to ${\bar B}_c$  in $\Xi_{bbq}$ Decay}

\author{Alexander K. Ridgway and Mark B. Wise}
\affiliation{Walter Burke Institute for Theoretical Physics,
California Institute of Technology, Pasadena, CA 91125}

\begin{abstract}

We estimate the branching ratio for  the inclusive decays $\Xi_{bbq} \rightarrow  {\bar B}_c^{(*)}+X_{c,s,q}$ to be approximately 1\%.  Our estimate is performed using non-relativistic potential quark model methods  that are appropriate if the bottom and charm quarks are  heavy compared to the strong interaction scale. Here the superscript $(*)$  denotes that we are summing over spin zero ${\bar B}_c$ and spin one ${\bar B}_c^*$ mesons and the subscript $q$ denotes a light quark. Our approach treats the two bottom quarks in the baryon $\Xi_{bbq}$ as a small color anti-triplet. This estimate for the inclusive branching ratio to ${\bar B}_c$  and  ${\bar B}^*_c$ mesons also holds for decays of the lowest lying $T_{bb{\bar q}{\bar q}}$ tetraquark states, provided they are stable against strong and electromagnetic decay.
\end{abstract}
\maketitle

\section{Introduction}

In 2017, the doubly charmed baryon $\Xi_{cc}^{++}$ (or in the notation used in this paper $\Xi_{ccu}$) was discovered at LHCb~\cite{Aaij:2017a}. It has been observed in the exclusive decay modes, $\Xi_{cc}^{++} \rightarrow  \Lambda_c^{+} K^{-} \pi^{+}\pi^{+}$ (the discovery mode) and $\Xi_{cc}^{++} \rightarrow \Xi_c^+\pi^+$~\cite{Aaij:2017b}. There is considerable interest in the detection of the analogous baryons containing two heavy bottom quarks $\Xi_{bbq}$, $q=u,d$, partly because it would be the first step to observing the tetraquark states, $T_{bb{\bar q}{\bar q}}$.  They are thought to be stable with respect to the strong and electromagnetic interactions with masses that are around 100-200 MeV below the ${ \bar B} _q {\bar B}_q$ threshold~\cite{Francis:2016hui, Eichten:2017ffp,Karliner:2017qjm}.

Recently, Gershon and Poluektov~\cite{Gershon:2018gda} proposed the inclusive decay mode $\Xi_{bbq} \rightarrow  {\bar B}_c +X_{{ c},s,q}$ as a potential discovery channel for the doubly bottom baryon $\Xi_{bbq}$ at the LHC. They made the clever observation  that $ {\bar B}_c$'s that do not point back to the  collision interaction point can only arise from the  weak decay of a hadron with two bottom quarks. They also note that  the decay chain ${\bar B}_c \rightarrow J/\psi \pi^- \rightarrow \mu^+ \mu^- \pi^-$ can be used to detect the ${\bar B}_c$ meson\footnote{See \cite{Qiao:2012hp} for a recent calculation of the branching ratio for ${\bar B}_c \rightarrow J/\psi \pi^- $. Their results imply that ${\rm Br}\left({\bar B}_c \rightarrow J/\psi \pi^- \rightarrow \mu^+ \mu^- \pi^-\right) \simeq 2 \times 10^{-4}$.}. Ordinary   ${\bar B}$ mesons  that do not point back to the collision point cannot be used for this purpose\footnote{We thank T. Gershon for pointing this out to us.} because they can arise from the weak decay of a long lived ${\bar B}_c$ meson (via the weak decay of the anti-charm quark). The branching ratio for  ${\bar B}_c$ decay  to ordinary ${\bar B}$ mesons is not expected to be  small and furthermore there will be many more ${\bar B}_c$'s produced at the interaction point by hadronization then there are baryons with two  bottom quarks.

 In this paper we make an estimate of the inclusive branching ratio,  ${ \rm Br}(\Xi_{bbq} \rightarrow  {\bar B}_c^{(*)} +X_{{ c},s,q})$.  Here the subscript $c,s,q$ denotes the flavor quantum numbers of the inclusive final state and the superscript $(*)$ denotes that we are summing over final state spin zero ${\bar B}_c$ and spin one ${\bar B}_c^*$ mesons. A ${\bar B}_c^*$ meson decays to a  ${\bar B}_c$ plus a photon, so decays to the spin one state always result in a ${\bar B}_c$ in the final state.
 
  Our method relies on treating both the bottom and charm quark as heavy compared to the scale of the non-perturbative strong interactions, $\Lambda_{QCD} \sim 200 {\rm MeV}$. In this limit, the two bottom quarks in the $\Xi_{bbq}$  form a small (compared with $1/\Lambda_{QCD}$) color anti-triplet diquark that we denote by $\Phi_{bb}$. Furthermore, the $\Phi_{bb}$ and ${\bar B}_c$  (and ${\bar B}^*_c$) can be treated as  non-relativistic bound states. The inclusive decay rate, $\Gamma(\Xi_{bbq} \rightarrow  {\bar B}_c^{(*)} +X_{{ c},s,q})$ is then modeled by $\Gamma(\Phi_{bb} \rightarrow  {\bar B}^{(*)}_c + c+s)$, with the light quark $q$ treated as a spectator.  This decay rate is easily converted into a branching ratio since the total decay rate of the $\Xi_{bbq}$ is approximately twice the $b$ quark decay rate\footnote{A more accurate estimate (which we will use) that applies the operator product expansion and heavy quark methods can be found in~\cite{Berezhnoy:2018bde}.}.
 
Our computation of  $\Gamma(\Xi_{bbq} \rightarrow  {\bar B}_c^{(*)} +X_{{ c},s,q})$ does not include decay products from an excited (radial  or  orbital)  ${\bar B}_c$ (or ${\bar B}^*_c$) mesons. We will calculate the decay rates to the first radially excited ${\bar B}_c$  and ${\bar B}^*_c$ mesons and show they are suppressed, and then argue that decays to the other excited states are suppressed as well.

Our calculation of the inclusive  decay rate of a $\Xi_{bbq}$ baryon to ${\bar B}_c$ and ${\bar B}^*_c$ mesons is similar to the calculation of the inclusive $B$ meson decay rate to $J/\Psi$~\cite{Wise:1979tp} . One important difference is that the baryon decay is not color suppressed. Another difference is that the baryon decay matrix element is proportional to an overlap of  wave-functions while the meson decay matrix element is proportional to the $ J/\psi $ wave function at the origin.

\section{The Decay Rate}

In this section, we outline the calculation of the  $\Phi_{bb} \rightarrow  {\bar B}_c + c+ s$ invariant matrix element ${\cal M}(\Phi_{{bb}}({\bf 0},\gamma) \rightarrow  {\bar B}_c({\bf k}) +  c({\bf p}_c,\alpha)+ s({\bf p}_s,\beta))$, where greek letters denote the color quantum numbers.  We perform the calculation in the rest frame of the decaying bottom diquark state $\Phi_{bb}$, which is a color anti-triplet and has spin one.  We assume that the relative momentum of the bound states are non-relativistic. The state vectors are then
\begin{align}
\label{states}
\ket{\bar{B}_c({\bf k},s,m_s)} &= \frac{\sqrt{2 E_{\bar{B}_c}(k)}}{\sqrt{3}}\int \frac{d^3 p}{(2\pi)^3}\tilde{\psi}_{\bar{B}_c}({\bf p})C^{s,m_s}_{s_1 s_2}\ket{b(\frac{m_b{\bf k}}{m_b+m_c}+{\bf p},\delta,s_1)\bar{c}(\frac{m_c{\bf k}}{m_b+m_c}-{\bf p},\delta,s_2)}\cr
\ket{\Phi_{bb}({\bf 0},\gamma,m)} &= \frac{1}{2}\sqrt{2 m_{\Phi_{bb}}}\int \frac{d^3 p}{(2\pi)^3}\tilde{\psi}_{\Phi_{bb}}({\bf p})\epsilon^{\gamma\alpha\beta}C^{1,m}_{s_1 s_2}\ket{b({\bf p},\alpha,s_1)b(-{\bf p},\beta,s_2)}
\end{align}
where repeated indices are summed over and the state $\ket{\bar{B}_c({\bf k},s,m_s)}$ corresponds to a $\bar{B}_c$ meson if $s=0$ and a $\Bar{B}^{*}_{c}$ meson if $s = 1$. The bound states have been normalized such that $\braket{\bar{B}_c({\bf k}_1,s_1,m_{s_1})|\bar{B}_c({\bf k}_2,s_2,m_{s_2})} = 2 E_{k_1} \delta^{s_1 s_2}\delta^{m_{s_{1}} m_{s_{2}}}(2\pi)^3\delta^3({\bf k}_1 - {\bf k}_2)$ and similarly for the $\Phi_{bb}$ state.  The state vectors on the right hand side of (\ref{states}) have no hidden normalization factors, and are just the appropriate creation operators acting on the vacuum.  The functions $\tilde{\psi}_{\bar{B}_c}({\bf p})$ and $\tilde{\psi}_{\Phi_{bb}}({\bf p})$ are the wavefunctions for the relative momentum of the quarks in the bound states and, in the non-relativistic limit, have support when ${\bf p}$ is much less than the masses of the bound quarks.

The weak Hamiltonian that induces the decay is\footnote{We neglect the contribution from operators induced by penguin type diagrams.}
\begin{equation}
H={4 G_F \over \sqrt{2}}V_{cs}^*V_{cb}\left[ C_1O_1 +C_2 O_2 \right]
\end{equation}
where 
\begin{align}
\label{operators}
O_1=\left[ {\bar c}_{\alpha} \gamma^{\mu}P_Lb_{\alpha} \right] \left[ {\bar s}_{\beta} \gamma_{\mu}P_Lc_{\beta} \right]~~~O_2=\left[ {\bar c}_{\beta} \gamma^{\mu}P_L b_{\alpha} \right] \left[ {\bar s}_{\alpha} \gamma_{\mu}P_L c_{\beta} \right].
\end{align}
The operators $O_{1,2}$ and coefficients $C_{1,2}$ are evaluated at a subtraction point equal to the $b$ quark mass. The invariant matrix element for the decay is then
\begin{align}
\label{matrixelement}
{\cal M}&={4G_F \over \sqrt{6}}V_{cs}^*V_{cb}(C_1-C_2) \int {d^3 p \over (2 \pi)^3}\frac{\sqrt{E_{{\bar B}_c}({k}) m_{\Phi_{bb}}}}{\sqrt{E_c(|{\bf p+k}|)E_b(p)}}{ \tilde{\psi}}_{\bar{B}_c}^*(|{\bf p}+\frac{m_b}{m_b+m_c}{\bf k} |) {\tilde{\psi}}_{\Phi_{bb}}  ({p})  \cr
& \times \epsilon_{\gamma \alpha \beta}  C^{(s,m_s)*} _{s_1,s_2'}C^{(1,m)}_{ s_1,s_2}\left[{\bar u^{(s)} }({\bf p_s},s_s)  \gamma^{\mu}P_L  v ^{(c)}({\bf p}+{\bf k},s_2') \right]\left[{\bar u^{(c)} }({\bf p_c},s_c)  \gamma_{\mu}P_L  u ^{(b)}({\bf p},s_2) \right]. 
\end{align}

In eq.~(\ref{matrixelement})  the $\tilde{\psi}_{\Phi_{bb}}(p)$ wavefunction restricts $p$ to be much less than $m_b$, so we can set $u ^{(b)}({\bf p},s_2) = u ^{(b)}({\bf 0},s_2)$ and $E_{b}(p) = m_b$.  In addition, the ${\bar B}_c$ wave function restricts $|{\bf p}+{m_b \over m_b+m_c} {\bf k} |$ to be much less than the charm quark mass, which means we can make the replacement  ${\bf p}+{\bf k} \rightarrow (m_c/ (m_b+m_c)){\bf k}$ in $E_c$ and $v^{(c)}$.  In the non-relativistic limit, the masses of the bound states are approximately equal to the sum of their constituent quark masses, which implies $E_c((m_c/(m_b+m_c))k) = (m_c/(m_c+m_b))E_{\bar{B}_c}(k)$.  After making these replacements, eq.~(\ref{matrixelement}) becomes
\begin{align}
\label{matrixelement1}
{\cal M}&\simeq {4G_F \over \sqrt{2}} V_{cs}^*V_{cb}(C_1-C_2 )\sqrt{{2(m_b+m_c)\over 3m_c}}\epsilon_{\gamma \alpha \beta}   C^{(s,m_s)*} _{s_1,s_2'}C^{(1,m)}_{ s_1,s_2}{\cal I} \left({m_b \over m_b+m_c}k\right)\cr
&\ \ \ \ \ \ \ \times[{\bar u^{(s)} }({\bf p_s},s_s)  \gamma^{\mu}P_L v ^{(c)}({m_c \over m_b+m_c}{\bf k},s_2')]  \left[{\bar u^{(c)} }({\bf p_c},s_c)  \gamma_{\mu}P_L  u ^{(b)}({\bf 0},s_2) \right] 
\end{align}
where
\begin{equation}
\label{formfactor}
{\cal I}(k)= \int {d^3 p \over (2 \pi)^3}{\tilde{\psi}}_{\bar{B}_c}^*\left(|{\bf p}+ {\bf k} | \right)  \tilde{\psi}_{\Phi_{bb}}  ({p}) = 4 \pi \int d r r^2  \psi_{\bar{B}_c}^*(r) \psi_{\Phi_{bb}}  (r) { {\rm sin}(kr) \over kr}.
\end{equation}
Note, the position space wavefunctions are normalized so that $\int |\psi_{\bar{B}_c/\Phi_{bb}}(r)|^2 d^3r = 1$. 

To determine the differential decay rate, we square the matrix element, average over initial spins and colors and sum over final spins and colors.  The spin sum involving the final state $\bar{B}_c$  spins is performed using the completeness relation,  
$\sum_{s,m_s} C_{s_a , s_b} ^{(s,m_s) } C_{{\bar s}_a, {\bar s}_b} ^{*(s,m_s)}=\delta_{ s_a  {\bar s}_a} \delta_{ s_b {\bar s}_b}$. For the spin average over the $\Phi_{bb}$ spin magnetic quantum numbers we note that, 
$\sum_{s_1}  C^{(1,1)}_{s_1,s_2}C^{*(1,1)}_{s_1, {\bar s}_2}+ C^{(1,-1)}_{s_1,s_2}C^{*(1,-1)}_{s_1,{\bar s}_2}=\delta_{s_2,{\bar s}_2} $.  Rotational invariance implies that the decay rate is independent of the magnetic quantum number for the total spin of $\Phi_{bb}$.  This means we can replace the average over its initial magnetic quantum numbers in the decay rate with the average over just the $m=-1$ and $m=1$ magnetic quantum numbers.  After integrating over the strange and charm momenta, the differential decay rate then becomes
\begin{eqnarray}
\label{differential decay}
{d\Gamma(\Xi_{bbq} \rightarrow  {\bar B}_c^{(*)}(k) +X_{{ c},s,q}) \over  dk}&&\simeq \left( G_F^2\over 3 \pi^3 \right)(C_1-C_2)^2|V_{cb}V_{cs}|^2|{\cal I}\left(m_bk/(m_b+m_c)\right)|^2  \nonumber \\
&&\ \ \ \ \ \ \ \times \ k^2{(m_{\Phi_{bb}}^2+m_{\bar{B}_c}^2-m_c^2-2 m_{\Phi_{bb}}E_{{\bar B}_c}(k))^2 \over (m_{\Phi_{bb}}^2+m_{\bar{B}_c}^2-2 m_{\Phi_{bb}}E_{{\bar B}_c}(k))}
\end{eqnarray}
where, as mentioned in the introduction, the superscript ${(*)}$ denotes that we are summing over the spin one and spin zero $\bar{B}_{c}$ mesons. 

\section{Numerical Results}

To evaluate the form factor ${\cal I}(k)$, we need to determine the wave functions $\psi_{\Phi_{bb}}(r)$ and $\psi_{\bar{B}_c}(r)$.  We do this by numerically solving the non-relativistic Schrodinger equation with the Cornell potentials,
\begin{equation}
\label{potential}
V_{\Phi_{bb}}(r) = - {2 \over 3}\left({0.3 \over r} \right)+{1 \over 2} (0.2 {\rm GeV}^2) r,~~~  V_{\bar{B}_c}(r)=-{4 \over 3}\left({0.4\over r}\right) +(0.2 {\rm GeV}^2) r.
\end{equation} 
The relative factor of $1/2$ between $V_{\Phi_{bb}}(r)$ and $V_{\bar{B}_{c}}(r)$ reflects the fact that the $\Phi_{bb}$ is a color anti-triplet while the $\bar{B}_c$ is a color singlet\footnote{Lattice studies indicate that the factor of one half should be extended to the non-perturbative linear part of the potential~\cite{Nakamura:2005hk}.}.  We took the string tension to be $0.2\ {\rm GeV}^2$ which fits the $b\bar{b}$ spectrum of bound states~\cite{ Bali:2000gf}.  In addition, we chose the strong fine structure constant to be $0.3$ and $0.4$ for $V_{\Phi_{bb}}(r)$ and $V_{\bar{B}_{c}}(r)$.
\begin{figure}
	\centering
	\includegraphics[width=4in]{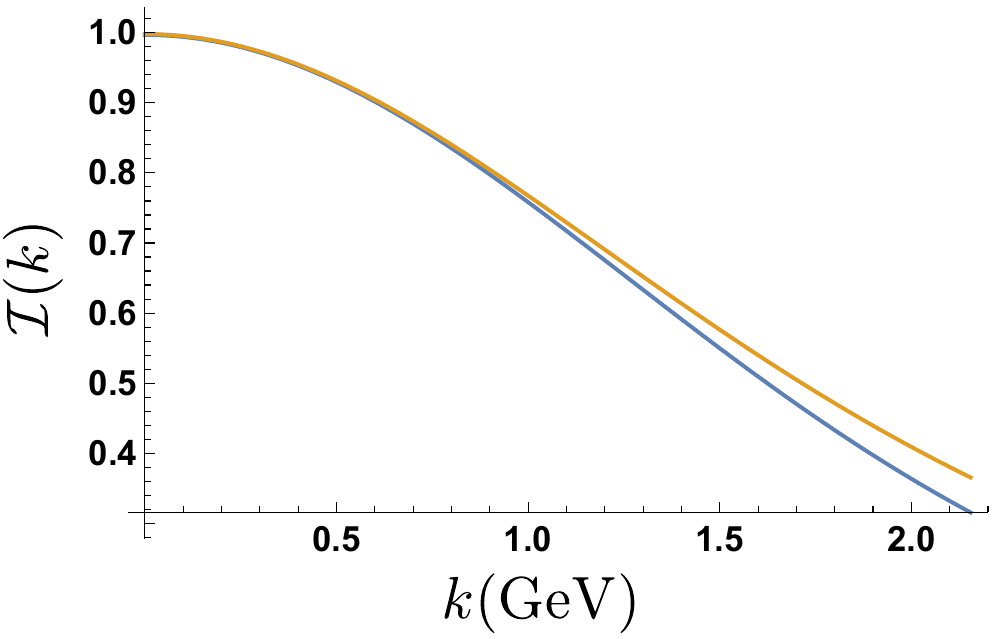}
	\caption{The form factor ${\cal I}(k)$ defined in (\ref{formfactor}) computed using the numerical ground state wave functions (blue) and the approximate ones (yellow) given in (\ref{analytic wavefunction})}
	\label{form factor fig}
\end{figure}	

The charm and bottom quark masses are taken to be  $1.5~{\rm GeV}$ and $4.5~{\rm GeV}$.  The form factor ${\cal I} (k)$ computed using the numerical ground state wavefunctions is plotted in Fig. \ref{form factor fig} over the range of $k$  allowed in the decay,
\begin{equation}
\label{phase space}
0<k< \left[  \left( (m_{\Phi_{bb}}^2+m_{{\bar B}_c}^2  -m_c^2)/ (2 m_{{\Phi}_{bb}}) \right)^2- m_{{\bar B}_c}^2 \right]^{1/2}.
\end{equation} 
The numerical solutions to the Schrodinger equation implies the radii squared of the ground state wavefunctions are $\langle r^2 \rangle_{\Phi_{bb}}=3.2 ~{\rm GeV}^{-2}$ and  $\langle r^2 \rangle_{\bar{B}_{c}}=2.8~ {\rm GeV}^{-2}$ .  It turns out that the Coulomb-like wave functions
\begin{align}
\label{analytic wavefunction}
\psi_{\Phi_{bb}}(r) &=\frac{1}{\sqrt{\pi}}\left(\frac{3}{\left< r^2 \right>_{\Phi_{bb}}}\right)^{3/4}{\rm Exp}\left({-\frac{\sqrt{3}r}{\sqrt{\langle r^2 \rangle_{\Phi_{bb}}}}}\right)\cr
\psi_{\bar{B}_{c}}(r) &= \frac{1}{\sqrt{\pi}}\left(\frac{3}{\left< r^2 \right>_{\Bar{B}_{c}}}\right)^{3/4}{\rm Exp}\left({-\frac{\sqrt{3}r}{\sqrt{\langle r^2 \rangle_{\Bar{B}_{c}}}}}\right)
\end{align}
are good approximations to the numerical ones.  Evaluating (\ref{formfactor}) using (\ref{analytic wavefunction}) gives the following simple analytic approximation to the form factor,
\begin{equation}
{\cal I} (k)= \left( { \langle r^2 \rangle_{\Phi_{bb}} ^{1/4}  \langle r^2 \rangle_{\bar{B}_{c}} ^{1/4}  \over    \langle r^2 \rangle_{\Phi_{bb}}^{1/2} +  \langle r^2 \rangle_{\bar{B}_{c}} ^{1/2}      }\right)^3{8 \over \left[  1+  \left( { \langle r^2 \rangle_{\Phi_{bb}}  \langle r^2 \rangle_{\bar{B}_{c}} /    (\langle r^2 \rangle_{\Phi_{bb}}^{1/2} + \langle r^2 \rangle_{\bar{B}_{c}} ^{1/2} )^2} \right)  \left( {k^2 / 3}\right)\right]^2}
\end{equation}
which is also plotted in Fig. \ref{form factor fig}.  

\begin{figure}
	\centering
	\includegraphics[width=4in]{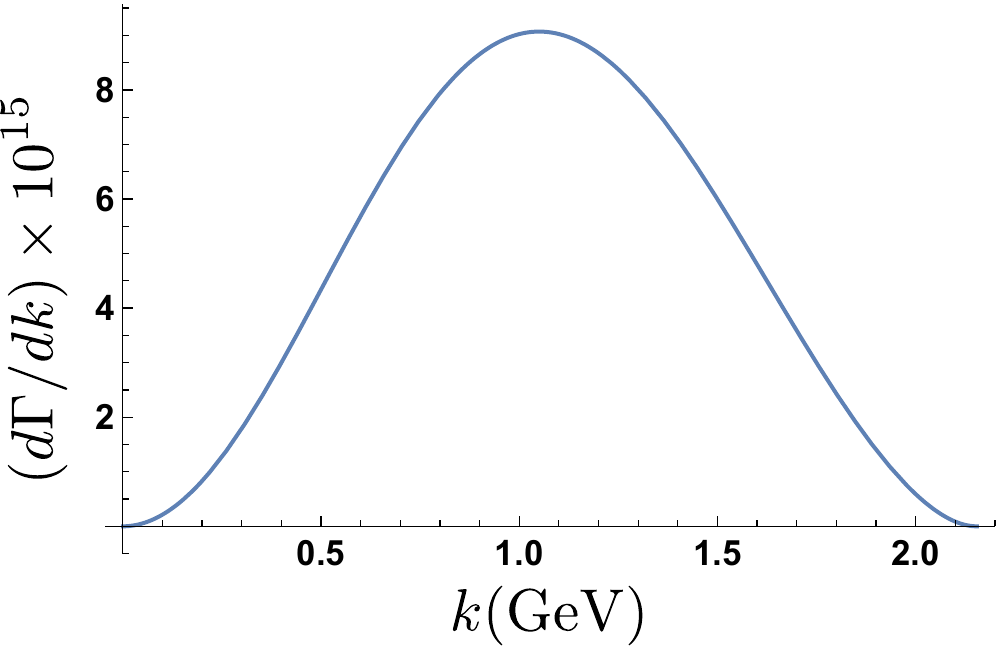}
	\caption{Derivative of the decay rate with respect to the momentum of the outgoing $\bar{B}_c$.}
	\label{decay rate plot}
\end{figure}

In Fig. \ref{decay rate plot} we plot $d\Gamma/dk$ obtained using the ground state numerical wavefunctions.  Integrating (\ref{differential decay}) over (\ref{phase space}), we find that the decay rate is 
\begin{align}
\Gamma(\Xi_{bbq} \rightarrow  {\bar B}_c^{(*)}(k) +X_{{ c},s,q})= 1.5\times10^{10}~ {\rm s}^{-1}.
\end{align}
Using a total lifetime for $\Xi_{bbq}$ of $0.5$ ps \cite{Berezhnoy:2018bde}, the branching ratio is 
\begin{align}
\label{branching ratio}
{\rm Br}(\Xi_{bbq} \rightarrow  {\bar B}_c^{(*)}(k) +X_{{ c},s,q}) \simeq 8 \times 10^{-3}.
\end{align}
This branching ratio leaves out $\bar{B}_c$'s that arise from the decay of radially and orbitally excited ${\bar B}_c$ and $ {\bar B}^{*}_c$ mesons.  We computed the decay rate to the first radially excited $\bar{B}_{c}$ state with zero orbital angular momentum and found the branching ratio to be $7.3\times 10^{-4}$, which is an order of magnitude smaller than the branching ratio to the ground state.

Decays to other radially excited $\bar{B}_{c}$ states will be suppressed as well.  The full Hamiltonian for the $\Phi_{bb}$ system, including the kinetic terms and potential from (\ref{potential}), is almost equal to half the full Hamiltonian for the $\bar{B}_{c}$ one.  This means the spatial wave functions for  the energy eigenstates of the two Hamiltonians are almost the same, which implies ${\cal I}(0) \simeq1$ for the ground state $\bar{B}_c$ mesons.  In addition, it implies that the overlap integral for decays to  radially excited $\bar{B}_{ c}$ and  $\bar{B}_{c}^*$ mesons will  satisfy ${\cal I}(0) \simeq 0$, which suppresses the branching ratio to these states.  Decays to orbitally excited $\bar{B}_c$ mesons are also suppressed.

A recent work on production rates for hadrons with two heavy quarks at the LHC \cite{Ali:2018xfq} estimates that  $\sigma (pp \rightarrow \Phi_{bb}+X)\simeq 15~{\rm nb}$.  Assuming most of the $\Phi_{bb}$ diquarks end up as $\Xi_{bbq}$ baryons\footnote{About $20\%$ end up as tetraquarks containing two bottom quarks.}  this implies that in an integrated luminosity of $10~ {\rm fb}^{-1} $ there are around $10^{8}$ $\Xi_{bbq} $ baryons. Our work then implies that the decays of these baryons produce around $10^{6}$ ${\bar B}_c$'s that do not point back to the interaction point.  About $10^2$ of them end up in the final state $\mu^{+}\mu^{-}\pi^{-}$, with the $\mu^{+}\mu^{-}$ arising from $J/\psi$ decay.

\section{Concluding Remarks}

We calculated the inclusive decay rate for $\Xi_{bbq} \rightarrow \bar{B}_{c} + X_{c,s,q}$ to be $1.5\times 10^{10}$ s$^{-1}$ (which implies ${\rm Br}(\Xi_{bbq} \rightarrow  {\bar B}_c^{(*)}(k) +X_{{ c},s,q}) \simeq 8\times 10^{-3}$).  The initial $bb$ system was treated as a tightly bound color anti-triplet diquark $\Phi_{bb}$ and we evaluated its decay rate to $\bar{B}_{c} + c + s$.  The Schrodinger equation was solved numerically to determine the non-relativistic wavefunctions for the $\Phi_{bb}$ and $\bar{B}_{c}$.  In reality, the relative momentum of the quarks in the $\bar{B}_{c}$ bound state is not truly non-relativistic.  In addition, we neglected the fact that the diquark initially exists in a hadron and interactions between the final $\bar{B}_{c}$, $c$ and $s$ states and the soft degrees of freedom in $\Xi_{bbq}$.  Despite these approximations, we expect our calculation of the decay rate to be correct at the factor of two level.

\acknowledgments
This work was supported by the  by
DOE Grant DE-SC0011632 . We are also grateful for the support provided by the Walter
Burke Institute for Theoretical Physics. We thank Zoltan Ligeti for informing us of the work
of Gershon and Poluektov. MBW also thanks Tim Gershon for discussions.

\end{document}